\documentclass[aps,twocolumn,superscriptaddress,prl,showpacs]{revtex4}

\usepackage{graphicx}

\begin{document}
\title{Bistability of the Nuclear Polarisation created through optical pumping in InGaAs Quantum Dots}
\author{P.-F.\ Braun}
\author{B.\ Urbaszek}
\email[Corresponding author : ]{urbaszek@insa-toulouse.fr}
\author{T.\ Amand}
\author{X.\ Marie}
\affiliation{LNMO, INSA, 135 avenue de Rangueil, 31077 Toulouse Cedex 4, France}
\author{O.\ Krebs}
\author{B.\ Eble}
\author{A.\ Lemaitre}
\author{P.\ Voisin}
\affiliation{Laboratoire de Photonique et Nanostructures, route de Nozay, 91460 Marcoussis, France}


\date{28 July 2006}

\begin{abstract}
We show that optical pumping of electron spins in individual InGaAs quantum dots leads to strong nuclear polarisation that we measure via the Overhauser shift $\delta_n$ in magneto-photoluminescence experiments between 0 and 4T.  We find a strongly non-monotonous dependence of $\delta_n$ on the applied magnetic field, with a maximum nuclear polarisation of about 40\% for intermediate magnetic fields. We observe that $\delta_n$ is larger for nuclear fields anti-parallel to the external field than in the parallel configuration. A bistability in the dependence of $\delta_n$  on the spin polarization of the optically injected electrons is found. All our findings are qualitatively understood with a model based on a simple perturbative approach. 
 \end{abstract}

\maketitle


\textbf{Introduction}

The spin of a single carrier confined to a semiconductor quantum dot (QD) can be manipulated either electrically or optically \cite{Zut}. Proposals for using single spins as building blocks for future memory or quantum computer architectures \cite{Awschbook} have been encouraged by the measurements of long spin relaxation times in the millisecond range \cite{Kroutvar1,oulton06}. In positively charged excitons the spin life- and coherence time of an electron in the ground state of a QD is determined by the interaction with the nuclear spins \cite{Merkulov02,Khaet1,Seme1,pif05} in the absence of electron-hole exchange. Following injection of electrons with a preferred spin direction, the electron spin can be imprinted on the nuclei in the dot via the hyperfine interaction \cite{Lampel68}. This dynamical polarisation of the coupled electron spin-nuclear spin system shows several surprising effects, such as strong internal magnetic fields in the order of Teslas \cite{GammonPRL,Bracker1}. There are many similarities between an electron in a quantum dot and electrons trapped by a donor in doped bulk semiconductors, for which the hyperfine interaction has been studied in great detail, for a review see \cite{opor}. Studies of spin polarised nuclei in GaAs show very long nuclear spin relaxation times \cite{opor,Paget82} and it would be interesting to transfer the spin information from the electrons onto the nuclei \cite{Ima2} which do interact far less with the lattice than electrons. 

Dynamical nuclear polarisation through optical pumping leads to the construction of an effective nuclear field $B_n$. In an applied magnetic field the Zeeman splitting of an electron is given by the contributions of the external field and $B_n$.  The contribution due to $B_n$ is the Overhauser shift (OHS) $\delta_n$ and can be measured by single dot photoluminescence (PL) spectroscopy in the Faraday geometry.
Work by Gammon and co-workers on GaAs interface fluctuation quantum dots on neutral and charged excitons show that dynamical polarisation is very effective in their samples with 60\% of the nuclei in the quantum dot polarised through optical pumping \cite{Gammon1,GammonPRL,Bracker1}. To evaluate the nuclear polarisation these authors compare the measured OHS with the maximum value theoretically obtainable for a nuclear polarisation of 100\%. In contrast, similar experiments on neutral (charged) excitons in self assembled InAlAs (InGaAs) quantum dots show a nuclear polarisation of only 6\% (10\%) \cite{Yokoi,Eble,Lai}. For self assembled InP dots effective nuclear fields $B_n$ of only a few mT have been observed \cite{Dzhioev1,Igna1} compared to effective fields of several Tesla in GaAs dots.
This is very surprising as for In containing compounds in principle large nuclear effects are expected due to the nuclear spin of 9/2 as compared to only 3/2 for both Ga and As , see table \ref{tab:table1}.
We show in this study that in order to achieve a substantial nuclear polarisation through optical pumping in self assembled InGaAs dots a careful analysis of the interdependence of applied magnetic field strength, optical pumping power and electron spin polarisation effects is necessary. We polarise up to 38\% of the nuclei in individual QDs corresponding to  $\delta_n=90\mu$eV.  We find a non-monotonous dependence of $\delta_n$ on the applied magnetic field, in contrast to GaAs QDs. We show that the magnetic field value, for which a maximum $\delta_n$ is measured, increases with optical pump power. We observe that $\delta_n$ is larger for nuclear fields anti-parallel to the external field than in the parallel configuration. Finally, we uncover a bistability in the dependence of $\delta_n$  on the spin polarization of the electron occupying the QD. All these findings can be modelled by calculating the stationary value of $\delta_n$ from a rate equation model. 

\textbf{Experimental details}

The sample contains the following layers, starting from the substrate: 200nm of p doped (Be) GaAs / 25nm of GaAs / InGaAs dots with a wetting layer grown in the Stranski-Krastanov mode / 30 nm of GaAs / 100nm Ga$_{0.7}$Al$_{0.3}$As / 20nm of GaAs. Placing a doped layer below the dots enables holes to tunnel into the dots. A low dot density between $10^8 cm^{-2}$ and $10^9 cm^{-2}$ adapted to single dot measurements has been obtained by choosing a nominal thickness of 1.7 mono-layers for the InAs layer deposition. The InGaAs quantum dots formed after Gallium and Indium interdiffusion contain typically 45\% Indium and 55\% Gallium,  their typical diameter is around 20nm and the height varies between 4 and 10 nm as determined by TEM measurements \cite {AL2004}.

The photoluminescence measurements on individual dots were carried out with a confocal microscope built around an Attocube nano-positioner placed in the centre of a superconducting magnet system at fields between 0 and 4T. The sample temperature in the variable temperature insert was kept at 1.5 K. The polarisation of the excitation as well as the detected signal was controlled with a Glan-Taylor polariser and a liquid crystal based wave plate. The optical signal was dispersed in a spectrometer and detected with Si-CCD Camera. The high signal to noise ratio single dot PL spectra were fitted with Lorentzian lineshapes that result in a spectral precision of our measurements of +/- 2.5 $\mu$eV. The sample was excited with a pulsed Ti-Sapphire laser with an 80 MHz repetition frequency. The laser spot size was about 1$\mu m^2$. The excitation pulses are circularly polarized $\sigma^+$. The luminescence intensity co-polarized ($I^+$) and counter-polarized ($I^-$) with the excitation laser are recorded. The circular polarization degree of the luminescence is then defined as $P_c = (I^+ - I^-)/ (I^+ + I^-)$. In the following the arrows $\uparrow$ , $\downarrow$ characterize the spin projection on the Oz growth axis of the electron ground states, the heavy hole spin is noted as $+\frac{3}{2}$ and $-\frac{3}{2}$.

We neglect here heavy-light hole mixing as the hole levels are well separated by strain. The excitation laser was set to a photon energy of 1.43 eV, optically injecting carriers into the low energy part of the wetting layer. We excite thus directly the heavy hole to electron transition in the wetting layer. Exciting the sample with a photon energy of 1.40 eV, which corresponds to intradot or crossed transitions (for example from a hole level in the wetting layer to an electron level in the dots, see reference \cite {Vasa2002}) did not alter the observed effects. We found that the circular polarisation degree of the emission, and therefore the average electron spin, remained basically unchanged when exciting at 1.40eV as compared to wetting layer excitation at 1.43eV.

We have done all the measurements on positively charged excitons X$^+$, with the resident holes originating from the Be doped layer. As a first step we can distinguish between neutral or charged excitons on the one hand and multiexciton complexes on the other hand by analyzing the PL intensity as a function of excitation power, with the excitons showing a sub-linear increase in the PL intensity with excitation power \cite{Aki2002}. To then distinguish between the neutral and the positively charged exciton, we analyze the circular polarisation degree. For PL transitions stemming from neutral excitons in InGaAs dots $P_c$ is only in the order of a few percent due to anisotropic exchange \cite{Mb2,Senes2,Tarta2,Lang2004,Eble}. Transitions with polarisation degrees $P_c$ in the order of 50\% at zero external field are attributed to the X$^+$ transition. The characteristic spectrum of the X$^{2+}$ transition consists of 3 lines, separated by 1meV and then 3meV going from high to low energy, as discussed in reference \cite{Ediger06}. In contrast, only one transition is expected for the X$^+$, allowing a clear experimental distinction between the two cases. The quantum dot transitions discussed in the remainder of this paper are thus identified as X$^+$. 

\textbf{The hyperfine interaction}

The presence of spin polarised electrons is essential for building up a nuclear field. This is the case during the radiative lifetime of the positively charged exciton X$^+$, where the holes form a spin singlet and the single electron interacts with the nuclei. The radiative lifetime of these pseudo-particles is about 1 ns \cite{pif05}. The analysis of the circular polarization of the $X^+$ luminescence in QDs following circularly polarized laser excitation will thus probe \emph{directly} the spin polarization of the electron as $\langle\hat{S}_z^e\rangle=-P_c/2$. 
The hyperfine interaction between an electron of spin $\hat{S}^e=\frac{1}{2}\hat{\sigma}^e$ confined to a quantum dot and $N$ nuclei is described by the Fermi contact Hamiltonian. 
\begin{equation}
\label{eq:eqHf}
\hat{H}_{hf} = \frac{\nu_0}{2}\sum_{j}A^j \vert\psi(\bar{r}_j)\vert^2 \left(2\hat{I}_z^j\hat{S}_z^e+ [\hat{I}_+^j\hat{S}_-^e+\hat{I}_-^j\hat{S}_+^e]\right)
\end{equation}
where $\nu_0$ is the two atom unit cell volume, $\bar{r}_j$ is the position of the nuclei $j$ with spin $\hat{I}^j$, the nuclear species are In, As and Ga. $A^j$ is the constant of the hyperfine interaction with the electron and $\psi(\bar{r})$ is the electron envelope function. Due to the $p$-symmetry of the periodic part of hole Bloch function the interaction of the hole via the Fermi contact Hamiltonian is neglected in the following \cite{Abra,Gr1}.

In this work we will focus on the dynamical polarisation in InGaAs quantum dots in an external magnetic field $B_z$ parallel to the sample growth direction, that is larger than both the local magnetic field $B_L$ (characterising the nuclear dipole-dipole interaction) and the Knight field $B_e$ (the effective magnetic field seen by the nuclei due to the presence of a spin polarised electron), which are in the order of mT. For the interesting effects when $B_z=B_L$ or $B_e$  see \cite{opor}, and for recent discussions in quantum dots \cite{Lai,oulton06,Aki2006}.

\begin{figure}
\includegraphics[width=0.4\textwidth]{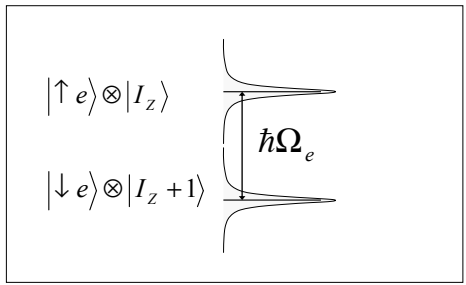}
\caption{\label{fig:fig0} Two states of the coupled electron-nuclear spin system are separated by an energy difference $\hbar\Omega_e$. The levels are broadened by $\hbar/\tau_c$ .
 }
\end{figure}

Introducing $\tilde{A}$ as the average of the hyperfine constants $A^j$ and assuming a strongly simplified, uniform electron wavefunction $\psi(\bar{r})=\sqrt{2/N\nu_0}$ over the involved nuclei equation \ref{eq:eqHf} simplifies to:
\begin{equation}
\label{eq:eqHf1}
\hat{H}_{hf} = \frac{2\tilde{A}}{N} \left(\hat{I}_z\hat{S}_z^e+ \frac{\hat{I}_+\hat{S}_-^e+\hat{I}_-\hat{S}_+^e}{2}\right)
\end{equation}

where $\hat{I}=\sum_{j=1}^{N}\hat{I}^j$. We take the first part of equation \ref{eq:eqHf1} and add the electron and nuclear Zeeman term to obtain: 

\begin{equation}
\label{eq:Hn}
\hat{H}_0=\hbar\gamma_nB_z\hat{I}_z/N+\hbar\Omega_e\hat{S}_z^e
\end{equation}

where $\hbar\Omega_e=g_e\mu_B(B_z+B_n)=\delta_z+\delta_n$. Here $\gamma_n$ is the nuclear gyromagnetic ratio, see table \ref{tab:table1} for the different nuclear species, $g_e$ is the longitudinal electron g-factor and $\mu_B$ is the Bohr magneton. Equation \ref{eq:Hn} gives rise to energy level splittings between the different nuclear and electron spin states. 
$\delta_n=2\tilde{A}\langle\hat{I}_z\rangle/N$ relates the Overhauser shift $\delta_n$ to the average nuclear polarisation. We can therefore access the average nuclear polarisation by measuring $\delta_n$.
For an order of magnitude calculation, we take the example of Indium (see table \ref{tab:table1}) and an electron g-factor of 1, and find  $(g_e\mu_B)/(\hbar\gamma_n)\simeq360$. We will thus neglect in the following the energy separation between the nuclear spin states. The second part of equation \ref{eq:eqHf1}, the spin flip-flop term: 

\begin{equation}
\label{eq:H1}
\hat{H}_1(t)=\frac{\tilde{A}}{N}(\hat{I}_+\hat{S}_-^e+\hat{I}_-\hat{S}_+^e)h_1(t)
\end{equation}

is a random perturbation between states split in energy by $\hbar\Omega_e$, see figure \ref{fig:fig0}. 
The function $h_1(t)$ is characterised by its mean value $\overline{h_1(t)}=f_e$ and a simple, normalised auto-correlation function $\overline{h_1(t)h_1^*(t+\tau)}=exp(-\frac{\vert\tau\vert}{\tau_c})$ with a correlation time $\tau_c$. The fraction of time the quantum dot contains an electron $f_e$ takes values between 0 and 1. For pulsed excitation $f_e$ can be evaluated as $f_e=\left(\frac{\alpha E_L}{h\nu}\right)\frac{\tau_{rad}}{T_L}$ where $\alpha$ is the absorption coefficient, as in the quantum well case dimensionless, $E_L$ is the energy per laser pulse,$\tau_{rad}$ is here the radiative lifetime of the X$^+$, $T_L$ is the laser repetition period. For cw excitation $f_e=\left(\frac{\alpha \overline{P}}{h\nu}\right)\tau_{rad}$ where $\overline{P}$ is the average excitation power. The formulas are approximations for small values of $f_e\leq0.1$ in the regime of linear absorption. The rate of nuclear polarisation will depend on the splitting $\hbar\Omega_e$ and the level broadening $\hbar/\tau_c$, see figure \ref{fig:fig0}. The upper limit of $\tau_c$ is given by $\tau_{rad}$ \cite{Text8}.

\begin{table}
\caption{\label{tab:table1} Comparison of the nuclear species, values taken from reference \cite{hand}}
\begin{ruledtabular}
\begin{tabular}{cccc}  
elements & In & Ga$(^1)$ & As  \\
\hline
\\
nuclear Spin $I$ & $\frac{9}{2}$ & $\frac{3}{2}$ & $\frac{3}{2}$ \\
hyperfine constant $A[\mu eV]$ & 56 & 42 & 46  \\
Electric quadrupole moment \\ 
$Q[10^{-24}cm^2]$ & 1.16 & 0.30 & 0.15 \\
$\hbar\gamma_n$ in units of $\mu_B$ & 2.75$\cdot10^{-3}$ & 1.14$\cdot10^{-3}$ & 0.72$\cdot10^{-3}$ \\
\hline
\\
$^1$ average of $^{69}$Ga and $^{71}$Ga 
\end{tabular}
\end{ruledtabular}
\end{table}

To access the average nuclear polarisation via the Overhausershift $\delta_n$ we chose the following experimental procedure: Depending on the circular polarisation direction of the pump beam, the effective nuclear field created is either parallel ($\vert B_{total} \vert=\vert B_z \vert + \vert B_n\vert $) or anti parallel ($\vert B_{total} \vert = \vert B_z \vert - \vert B_n \vert$) to the applied magnetic field $B_z$. To measure the OHS in both cases six PL spectra are needed. We measure the intensity of the $\sigma^+$ and $\sigma^-$ polarised PL intensity for three different excitation polarisations: $\sigma^+$ and $\sigma^-$ and linear. The typical accumulation time for one spectrum was 30 seconds. It should be noted that, as we changed the excitation polarisation between measurements, the order in which we recorded the spectra did not change the OHS measured. For instance, the Zeeman splitting following linearly polarised excitation did not change if we polarised the nuclei in a preceding measurement or not.

When a fraction of the nuclei in a quantum dot is optically polarised through $\hat{H}_{1}$, the electron will experience a total magnetic field $B_{total}$, whereas the hole will only experience the external field $B_z$.
Following excitation with linearly polarised light, injecting either a spin down or a spin up electron, the emitted PL energy will be shifted by either $E^{lin}_{+1}$ for $\sigma^+$ polarised PL, or $E^{lin}_{-1}$ for $\sigma^-$ polarised PL. 

\begin{equation}
\label{eq:Elin}
E^{lin}_{\pm 1}=\mp g_e\mu_B B_z/2 \pm g_h\mu_B B_z3/2
\end{equation}

where $g_{e(h)}$ is the longitudinal electron (hole) g-factor. For excitation with linear polarisation $B_n$ is zero as the average electron spin polarisation is zero. This gives rise to the following Zeeman splitting:

\begin{equation}
\label{eq:equZl}
\Delta Z(lin)=g_e\mu_B(-B_z)+3g_h\mu_BB_z 
\end{equation}

Following excitation with $\sigma^+$ polarised light, injecting a spin down electron $(\downarrow)$, the emitted PL energy will be shifted by either $E^p_{+1}$ for $\sigma^+$ polarised PL, or $E^p_{-1}$ for $\sigma^-$ polarised PL.

\begin{equation}
\label{eq:Epp}
E^p_{\pm 1}=\mp g_e\mu_B[B_z+B_n(\sigma^+)]/2 \pm g_h\mu_B B_z3/2
\end{equation}

In both cases the nuclear field is the same as denoted by $B_n(\sigma^+)$. This results in a Zeeman splitting that is different from the linear excitation case due to the presence of the nuclear field.

\begin{equation}
\label{eq:equZsp}
\Delta Z(\sigma^+)=\Delta Z(lin)-g_e\mu_B(B_n(\sigma^+))
\end{equation}

By subracting (\ref{eq:equZl}) from (\ref{eq:equZsp}) we obtain the OHS for $\sigma^+$ excitation; an identical argumentation leads to the OHS for $\sigma^-$ excitation.

\small
\begin{eqnarray} 
\label{equOHS}
\delta_n(\sigma^+)=\Delta Z(\sigma^+)-\Delta Z(lin)=-g_e\mu_BB_n(\sigma^+) \\
\delta_n(\sigma^-)=\Delta Z(\sigma^-)-\Delta Z(lin)=-g_e\mu_BB_n(\sigma^-) 
\end{eqnarray}\\
\normalsize 

It is important to note that the absolute values of $\delta_n(\sigma^+)$ and $\delta_n(\sigma^-)$ are only equal if $\vert B_n(\sigma^+)\vert = \vert B_n(\sigma^-)\vert$. Our experiments show in the following, that this is not the case.

\textbf{Experimental results and discussion}

The figures discussed in the following show phenomena that are typical for the analysis of several tenth of dots. The magnetic field, excitation power and electron spin polarisation dependence of the OHS was qualitatively the same for all the dots investigated. The maximum of the OHS varied between 60 and 90$\mu$eV from dot to dot, for reasons that are detailed below. Please note that as we plot the Zeeman splitting as a function of applied field and not the transition energy, no diamagnetic shift can be seen in our figures as it cancels out and is therefore neglected in the following discussion.

\begin{figure}
\includegraphics[width=0.47\textwidth]{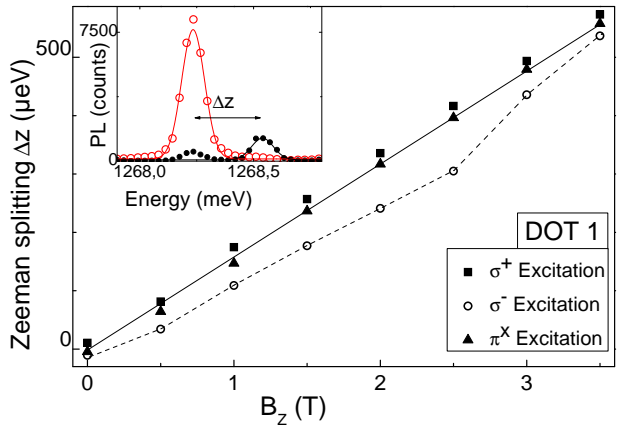}
\caption{\label{fig:fig1} The Zeeman splitting $\Delta Z$ is measured for an individual quantum dot for three different excitation laser polarisations: $\sigma^+$ (solid squares),$\sigma^-$ (hollow circles) and linear (solid triangles) for an excitation power of P=$8.7\mu W$. The dotted line is a guide to the eye. Inset: Single dot PL at 4T showing a clear Zeeman splitting between $\sigma^+$ PL (solid circles) and $\sigma^-$ PL (hollow circles) following $\sigma^-$ excitation.
 }
\end{figure}
Figure \ref{fig:fig1} shows the Zeeman splitting for a single quantum dot following linearly and circularly polarised excitation. We have verified experimentally that changing the direction of $S_z^e$ has the same effect as changing the field direction from $+B_z$ to $-B_z$. In this work we only show measurements for the same direction of $B_z$ for changing $S_z^e$. The Zeeman splitting following linear excitation is due to the external applied fields and grows, as expected, linearly. In contrast, when exciting the sample with $\sigma^+(\sigma^-)$ polarised light, the Zeeman splitting increases (decreases). This has been observed in GaAs interface fluctuation dots \cite{GammonPRL}.

\begin{figure}
\includegraphics[width=0.47\textwidth]{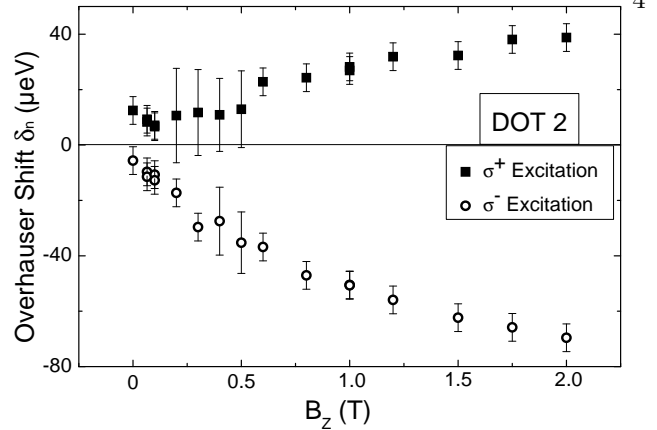}
\caption{\label{fig:fig2} The Overhauser shift $\delta_n$ is measured for magnetic fields up to 2 Tesla following $\sigma^+$ (solid squares) and $\sigma^-$ (hollow circles) excitation. For field values where the fitting procedure was more complex due to a small Zeeman splitting and a finite rejection ratio of the polarisation optics, we have increased the error bars accordingly.}
\end{figure}

Figure \ref{fig:fig2}  shows the values of the OHS for fields up to 2 Tesla in greater detail for another dot. We note: 

(i) $\delta_n$ does not reach the same absolute values for $\sigma^+$ and $\sigma^-$ excitation. The total magnetic field seen by the electron is smaller in the case of $\sigma^-$ excitation than in the case of $\sigma^+$ excitation, as in the case of $\sigma^-$ excitation the external magnetic field and the nuclear field are anti-parallel . This can be seen directly from the smaller Zeeman splitting in figure \ref{fig:fig1}. As the spin flip of the electron means going from one Zeeman level to the other, separated by $\hbar\Omega_e$, nuclear polarisation due to spin flip-flop is more efficient when the total magnetic field is small, here in the case of $\sigma^-$ excitation. This would explain the larger absolute value of the OHS measured for $\sigma^-$ excitation. To indicate why this has not been observed for GaAs dots the relative magnitude of the electron Zeeman splitting $\delta_z$ and $\delta_n$, that add up to $\hbar\Omega_e=\delta_z+\delta_n$, have to be taken into account, see table \ref{tab:table2}. In GaAs dots, $\delta_z$ is much smaller than in InGaAs due to the small electron g-factor , whereas $\delta_n$ is larger in GaAs than in InGaAs. The total energy splitting $\hbar\Omega_e$ for $\sigma^+$ ($\delta_z$ and $\delta_n$ have the same sign) and $\sigma^-$ ($\delta_z$ and $\delta_n$ have opposite signs) excitation are therefore not very different in GaAs, leading to a $\delta_n$ that is symmetrical within the experimental uncertainties \cite{Bracker1}.

\begin{table}
\caption{\label{tab:table2} Comparison of the electron Zeeman splitting $\delta_z=g_e\mu_BB_z$ and the maximum Overhauser shift $\delta_n$ experimentally observed, values taken from this work for InGaAs and reference \cite{GammonPRL} for GaAs.}
\begin{ruledtabular}
\begin{tabular}{ccc}  
at $B_z=1T$ & InGaAs dots & GaAs dots \\
\hline
electron g factor & 0.6 & 0.2 \\
$\delta_z [\mu eV]$ & 35 $\pm 2.5$ & 12  \\
 $\delta_n [\mu eV]$ & 40 $\pm 10$& 90 \\

\end{tabular}
\end{ruledtabular}
\end{table}

(ii) The OHS does increase with increasing magnetic field up to about 2T. For small fields in the order of mT this can be understood in terms of the suppression of the dipole-dipole interaction between nuclei, that leads to nuclear spin relaxation \cite{opor}. Once this mechanism is supressed it is surprising to find a further increase of the OHS with the applied magnetic field. In reference \cite{GammonPRL} for pure GaAs dots the OHS does not change measurably between 0.2T to 3.5T. We interpret the increase in OHS observed in our experiment as a function of the magnetic field as a gradual suppression of one or several nuclear depolarisation mechanisms, induced, for instance, by nuclear quadrupole coupling. Self-assembled, nominally pure InAs QDs are in reality InGaAs QDs due to In and Ga interdiffusion. This will induce local lattice distortion and strain \cite{Zunger2004}, giving rise to electric field gradients that lead to quadrupole coupling between the nuclear spin states \cite{Deng,Igna1}. It has already been suggested that nuclear quadrupole coupling is responsible for the small nuclear fields in self assembled InP dots \cite{Dzhioev1,Igna1}. This coupling, which exists for all nuclei with $I\ge3/2$ in a lattice with a symmetry that is lower than cubic, does not induce any nuclear spin relaxation itself, but it mixes the different nuclear spin states. Transitions are then possible due to fluctuations of the occupation of the dot by an electron. The degree of the mixing depends on the energy separation i.e. the Zeeman splitting between the different nuclear spin states. Increasing the external field will increase the nuclear Zeeman splitting and hence decouple the nuclear spin states, as a simple perturbation treatment shows \cite{Deng}. This would explain why our measurement show a steady increase in the OHS with the applied external field. As strain gradients and alloy composition vary from dot to dot on a microscopic scale, the maximum OHS will not reach the same value for every dot. The quadrupole coupling in GaAs interface fluctuation dots is certainly much weaker as the dots consist of a binary compound and are not strained, which could explain that the OHS does not measureably depend on the applied magnetic field for these dots. The case of InAlAs/AlGaAs dots studied in reference \cite{Yokoi} should be similar to our case, although the authors mention no dependence of the OHS on the magnetic field. But a closer look at their experimental data hints at variations in the OHS with a maximum at about 2 Tesla. Note that the Zeeman splitting at zero applied field is non-zero in figures \ref{fig:fig1} and \ref{fig:fig2}. When the superconducting magnet is switched off, we do measure a residual magnetic field in the order of $10^{-4}$ Tesla. When optically injecting electrons into the dot it is possible that a combination of the residual field and the Knight field are enough to suppress the nuclear dipole-dipole interaction and to allow dynamical nuclear polarisation. To study the effects of the Knight field in more detail, which is beyond the scope of this paper, the external fields have to be compensated by Helmholtz coils as in \cite {Lai, oulton06}.

\begin{figure}
\includegraphics[width=0.47\textwidth]{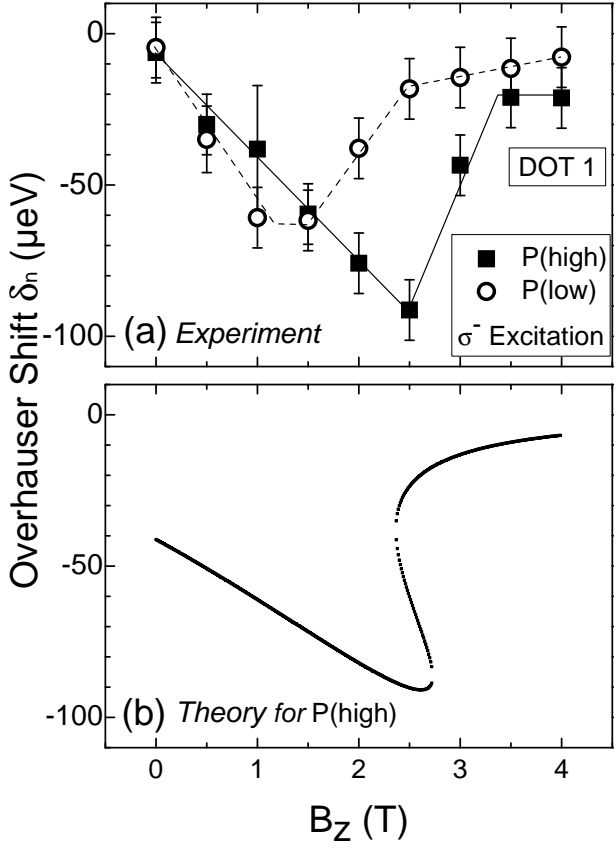}
\caption{\label{fig:fig3} (a) The Overhauser shift $\delta_n$  for $\sigma^-$ excitation is plotted as a function of the applied magnetic field for two different laser pump powers $P_{high}=8.7\mu W$ (solid squares) and $P_{low}=3.1\mu W$ (hollow squares). (b) Simulation of the experiment for $P_{high}$ using equation \ref{eq:nucpol} with the fitting parameters $2\tilde{A}\tilde{Q}$=1.3meV, $\tilde{T_d}$=2.5ms, $\tau_c$=19ps, $f_e$=0.05, $P_c$=0.6 and $g_e$=0.6 .} 
\end{figure}

So far we have discussed the \emph{increase} of the OHS with the applied magnetic field, but our measurements for values above 2 Tesla show an abrupt \emph{decrease} of the OHS for $\vert B_{total} \vert = \vert B_z \vert - \vert B_n \vert$ from a maximum absolute value of 90$\mu$eV at 2.5T to 20$\mu$eV at 3.5T, as can be seen in figure \ref{fig:fig1}, and even more clearly in figure \ref{fig:fig3}a. For a high excitation power $P_{high}=8.7\mu W$ we observe a maximum OHS at 2.5 Tesla. Taking a dot composition of In$_{0.45}$Ga$_{0.55}$As, the maximum Overhauser shift for a nuclear polarisation of 100\% is about 236$\mu eV$, the measured $\delta_n=90\mu eV$ correspond therefore to a nuclear polarisation of 38\%. For a lower excitation power of $P_{low}=3.1\mu W$ we observe a similar behaviour: the OHS increases with increasing applied magnetic field, reaches a maximum and then decreases. The maximum OHS measured for low excitation power is obtained at around 1.5 Tesla, see figure \ref{fig:fig3}.

\begin{figure}
\includegraphics[width=0.47\textwidth]{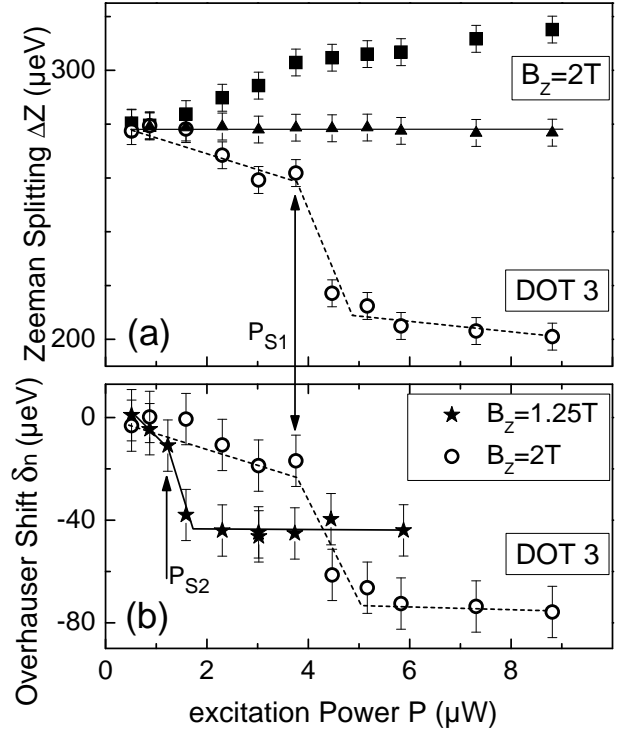}
\caption{\label{fig:fig4} (a) The Zeeman splitting $\Delta Z$ is measured for an individual quantum dot for three different excitation laser polarisations: $\sigma^+$ (solid squares),$\sigma^-$ (hollow circles) and linear (solid triangles) as a function of excitation power at a \emph{constant} field of $B_z=2T$. (b) The Overhauser shift $\delta_n$ for $\sigma^-$ excitation  at a field of $B_z=2T$ (hollow circles) and $B_z=1.25T$ (solid stars) as a function of excitation power.}
\end{figure}

The complicated magnetic field - excitation power interdependence is clarified in a second experiment. Figure \ref{fig:fig4}a shows the Zeeman splitting measured at a \emph{constant} magnetic field for \emph{different} pump powers. Not only the magnetic field dependence but also the power dependence is very different when comparing the case of $\vert B_{total} \vert = \vert B_z \vert - \vert B_n \vert$ and $\vert B_{total} \vert = \vert B_z \vert + \vert B_n \vert$. For the measurements at 2 Tesla for $\vert B_{total} \vert = \vert B_z \vert - \vert B_n \vert$ we observe a sudden increase in the absolute value of the OHS at an excitation power of $P_{S1}\simeq 3.7\mu W$ in figure \ref{fig:fig4}b. Increasing the pump power further does not increase the OHS significantly. For the measurements carried out at 1.25T we observe qualitatively the same behaviour, but with a threshold at a much lower pump power $P_{S2}\simeq 1.3\mu W$.  This can be understood when taking into account both the magnetic field (figure \ref{fig:fig3}a) and the power dependence (figure \ref{fig:fig4}): the magnetic field value $B_z$ that gives the maximum OHS depends on the pump power. Experiments carried out with a cw Ti-Sapphire Laser with a photon energy of 1.43eV gave very similar results, the main difference being the peak power of the two sources, which changes $f_e$.

\begin{figure}
\includegraphics[width=0.47\textwidth]{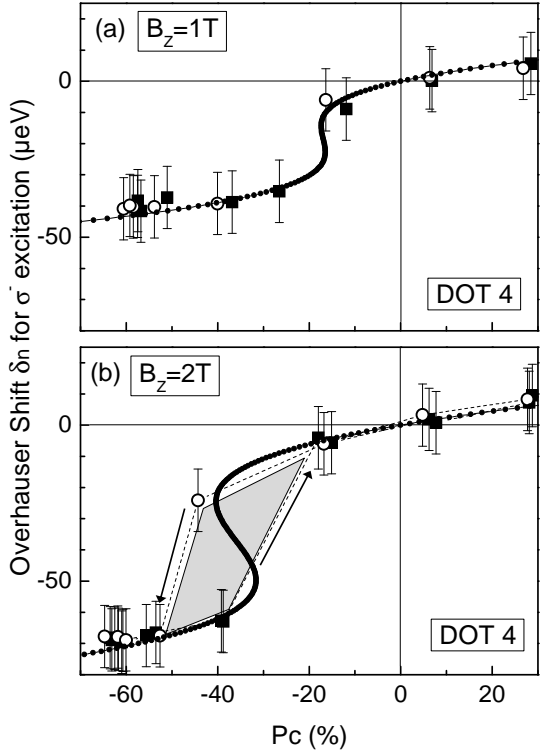}
\caption{\label{fig:fig5} (a) The Overhauser shift $\delta_n$ for $\sigma^-$ excitation  at a field of $B_z$=1T is plotted as a function of the measured PL polarisation $P_c$, going first from positive to negative $P_c$ (hollow circles) and then back from negative to positive $P_c$ (solid squares) for an excitation power of 8.7$\mu$W. A fit of the data (dotted line) using equation \ref{eq:nucpol} and $2\tilde{A}\tilde{Q}$=1.3meV, $\tilde{T_d}$=0.98ms, $\tau_c$=50ps, $f_e$=0.05, and $g_e$=0.48 hints at a bistability. Note that we do not show the full range for positive $P_c$ as $\delta_n$ changes very little between $P_c=30\%$ and 60\%.(b) The same measurement as (a), but at $B_z$=2T. The dashed lines are a guide to the eye. The two experimental curves do not perfectly overlap, this bistability (shaded region) effect is discussed in the text. For the fit (dotted line) only the values of  $\tilde{T_d}$=1.75ms and $\tau_c$=31ps have been changed as compared to (a).}
\end{figure}

Assuming a uniform nuclear polarisation and a high nuclear spin temperature, the nuclear polarisation rate in our system can be described by time dependent perturbation theory up to second order similar to reference \cite{Eble} and initially based on reference  \cite{Abra}.

\begin{equation}
\label{eq:rate}
\frac{d\langle\hat{I}_z\rangle}{dt}=-\frac{1}{T_{1e}}(\langle\hat{I}_z\rangle-\tilde{Q}\langle\hat{S}^e_z\rangle)-
\frac{\langle\hat{I}_z\rangle}{\tilde{T_d}}
\end{equation}

where $\tilde{Q}=\sum_{j}x_{j}\frac{I^{j}(I^{j}+1}{S(S+1)}$ and j=As,Ga

and

\begin{equation}
\label{eq:Te}
\frac{1}{T_{1e}}=\left(\frac{\tilde{A}}{N\hbar}\right)^2\frac{2f_e\tau_c}{1+(\Omega_e\tau_c)^2}
\end{equation}

In this expression we have assumed for simplicity that $\tilde{T_d}$ is an average nuclear decay constant, independent of the nuclear species. Here $\tilde{A}/(N\hbar)$ is the precession frequency of a nuclear spin in the Knight field of an electron, see the chapter of Dyakonov and Perel in \cite{opor}. This leads to an implicit expression for the equilibrium nuclear polarisation 

\begin{equation}
\label{eq:nucpol}
\langle\hat{I}_z\rangle=\tilde{Q}\langle \hat{S}_z^e \rangle\frac{\tilde{T_d}}{\tilde{T_d}+T_{1e}(\langle\hat{I}_z\rangle)}
\end{equation}

and hence for $\delta_n$.



The idea behind this model is to qualitatively explain the surprising power and magnetic field dependence of $\delta_n$. Equation \ref{eq:nucpol} is asymmetric in $B_z$ or $S_z$, different solutions are thus expected for the parallel and anti parallel case. It is useful to note that equation \ref{eq:nucpol} has only one real solution when $g_e\mu_B B_z$ and $\delta_n$ have the same sign, but may have up to three solutions when the signs are opposite, depending on the experimental conditions. This remarkable feature of equation \ref{eq:nucpol} does allow in principle the existence of bistability effects, predicted in the chapter of Dyakonov and Perel in \cite{opor} and by Artemova et al in reference \cite{Arte}, as can be seen in the simulations in figure \ref{fig:fig3}b and \ref{fig:fig5}a and \ref{fig:fig5}b. As for all the results discussed up to here we changed excitation polarisation for every experimental point, it is impossible to observe any bistability effects in the power or magnetic field dependence of the OHS, as very recently discussed by \cite{Tarta1,Imapriv}. Figure \ref{fig:fig3}b shows a fit of the experimentally observed magnetic field dependence with N=10000, the other parameters are given in the figure caption. The maximum at around 2.5 T and the sudden drop in the absolute value of the OHS are well reproduced, whereas the fit is less good for fields below 1 T. For a better fit a dependence of the nuclear decay constant $\tilde{T}_d$ on the applied magnetic field through the quadrupole coupling has to be included. 

In the next part we present the results obtained with a different experimental procedure. The nuclei get polarised by optically injected electrons with a preferred spin polarisation. We are able to vary the average spin $\langle\hat{S}_z^e\rangle$ of the injected electron by changing the relative retardation of the excitation beam continuously from $\lambda/4$ to $3\lambda/4$ by adjusting the voltage applied to the liquid crystal retarders. Please note that as we record two spectra, $\sigma^+$ and $\sigma^-$ polarised, for any given excitation polarisation, we measure both the OHS and the circular polarisation degree $P_c$. As $\langle\hat{S}_z^e\rangle=-P_c/2$ we are able to measure the OHS as a function of the average electron spin $\langle\hat{S}_z^e\rangle$ in the dot \cite{Eble}. For this experiment the order in which measurements were performed is important, as indicated by the arrows in figure \ref{fig:fig5}b. Looking at the third quadrant of figure \ref{fig:fig5}b we see that when increasing the absolute value of $P_c$ we observe a very abrupt increase in the OHS between the values of $P_c=-44\%$ to -53\%. Once the maximum of $\vert P_c \vert$ is reached, following an ideal $\sigma^-$ excitation, we decrease the absolute value of $P_c$ (by increasing the ellipticity of the excitation beam) and observe again a sudden change of the OHS, but now between the values of $P_c=-20\%$ to -38\%. That means the nuclear polarisation remained stable once it was created. This experiment demonstrates directly the dependence of the nuclear field created on the spin of the electron occupying the quantum dot. The experimentally observed bistability is very well reproduced by our model, as can be seen in the figure. The effect is far more pronounced at applied fields of 2T, see figure \ref{fig:fig5}b, for which the nuclear polarisation achieved is higher than for 1T, see figure \ref{fig:fig5}a. Note that a similar experiment performed in reference \cite{Yokoi} did not show any bistability effects or sudden jumps as it was carried out at 5T, which is according to our data a magnetic field value that is too high to observe a substantial nuclear polarisation.

\textbf{Conclusion}

We have presented a detailed experimental study of the dynamical polarization of nuclear spins through optical pumping in InGaAs quantum dots. To obtain a strong nuclear polarisation the applied magnetic field strength, the optical pump power and the optically created electron spin polarisation have to be optimised. We have demonstrated that on the one hand $B_z$ should not cause a too large electron Zeeman splitting, that makes spin flip-flop processes necessary to build up a nuclear polarisation too costly in energy. On the other hand we believe that $B_z$ has to be strong enough to decouple the nuclear spin states that are mixed by the quadrupolar interaction. We find experimentally that fields between 1.5 and 2.5 T fulfil both criteria for the excitation power range investigated. The higher the fraction of time $f_e$ the quantum dots contains an electron, the easier it is to dynamically polarise the nuclei in the dot. As $f_e$ increases for both cw and pulsed excitation with the laser power (see above) the Overhauser shift $\delta_n$ increases as well, a trend qualitatively reported in the literature \cite{Bracker1}. We find for the case of $\vert B_{total} \vert = \vert B_z \vert - \vert B_n \vert$ a threshold like increase of the absolute value of $\delta_n$ with power that saturates. The value of  $\delta_n$ that is reached at saturation in the power dependent experiments is highest in an applied field $B_z$ of about 2T. For $\vert B_{total} \vert = \vert B_z \vert + \vert B_n \vert$ the increase is less abrupt and the absolute value of $\delta_n$ reached is lower. Through optical pumping of self assembled InGaAs dots we find a nuclear polarisation of 38\%. The dependence of the nuclear polarisation on the magnetic field and the electron polarisation are well reproduced by our model. To determine if all the nuclear species present in the dot contribute to this polarisation nano-NMR measurements can be performed \cite{Gammon1}. The spectra that give the largest Overhauser shift  $\delta_n$ also show the highest electron spin polarisation. 
To reach the maximum $\delta_n$ a certain threshold value of $\langle\hat{S}_z^e\rangle$ is necessary. This threshold value is larger when going from low to high spin polarisation, than when going from high to low spin polarisation. This bistable behaviour means that once the nuclear field is created via a certain electron spin polarisation, it can be maintained with a lower electron spin polarisation. Our experiments seem to confirm that bistability is a general property of dynamical nuclear polarisation as it has previously been observed by magneto-PL in GaAs/AlGaAs (100) quantum wells \cite {Kal92} and more recently in time resolved Faraday rotation measurements in GaAs/AlGaAs (110) quantum wells \cite{San2004}. The bistable behaviour can be used to drastically change the effective nuclear field $B_n$ through a slight variation of an external parameter, here the generated electron polarisation. We show in our work a high level of control over the dynamical nuclear polarisation in an individual quantum dot. And it is this level of control that can be useful for future attempts of manipulating the state of the coupled electron-nuclear spin system in a single nano-object.

\textbf{Acknowledgements:} We thank Atac Imamoglu, Patrick Maletinsky and Vladimir Kalevich for fruitful discussions and Julian Carrey, Reasmey Tan and Marc Respaud for technical assistance. P.-F.B. acknowledges financial support from the FSE.
\\

\bibliographystyle{apsrev}

\begin{thebibliography}{35}
\expandafter\ifx\csname natexlab\endcsname\relax\def\natexlab#1{#1}\fi
\expandafter\ifx\csname bibnamefont\endcsname\relax
  \def\bibnamefont#1{#1}\fi
\expandafter\ifx\csname bibfnamefont\endcsname\relax
  \def\bibfnamefont#1{#1}\fi
\expandafter\ifx\csname citenamefont\endcsname\relax
  \def\citenamefont#1{#1}\fi
\expandafter\ifx\csname url\endcsname\relax
  \def\url#1{\texttt{#1}}\fi
\expandafter\ifx\csname urlprefix\endcsname\relax\def\urlprefix{URL }\fi
\providecommand{\bibinfo}[2]{#2}
\providecommand{\eprint}[2][]{\url{#2}}

\bibitem[{\citenamefont{Zutic et~al.}(2004)\citenamefont{Zutic, Fabian, and
  Sarma}}]{Zut}
\bibinfo{author}{\bibfnamefont{I.}~\bibnamefont{Zutic}},
  \bibinfo{author}{\bibfnamefont{J.}~\bibnamefont{Fabian}}, \bibnamefont{and}
  \bibinfo{author}{\bibfnamefont{S.~D.} \bibnamefont{Sarma}},
  \bibinfo{journal}{Rev. Mod. Phys.} \textbf{\bibinfo{volume}{76}},
  \bibinfo{pages}{323} (\bibinfo{year}{2004}).

\bibitem[{Aws()}]{Awschbook}
\bibinfo{note}{\emph{Semiconductor Spintronics and Quantum Computation}, edited
  by D.D. Awschalom, D. Loss and N. Samarth, NanoScience and Technology,
  (Springer, Berlin, 2002).}

\bibitem[{\citenamefont{Kroutvar et~al.}(2004)\citenamefont{Kroutvar, Ducommun,
  Heiss, Bichler, Schuh, Abstreiter, and Finley}}]{Kroutvar1}
\bibinfo{author}{\bibfnamefont{M.}~\bibnamefont{Kroutvar}},
  \bibinfo{author}{\bibfnamefont{Y.}~\bibnamefont{Ducommun}},
  \bibinfo{author}{\bibfnamefont{D.}~\bibnamefont{Heiss}},
  \bibinfo{author}{\bibfnamefont{M.}~\bibnamefont{Bichler}},
  \bibinfo{author}{\bibfnamefont{D.}~\bibnamefont{Schuh}},
  \bibinfo{author}{\bibfnamefont{G.}~\bibnamefont{Abstreiter}},
  \bibnamefont{and} \bibinfo{author}{\bibfnamefont{J.~J.}
  \bibnamefont{Finley}}, \bibinfo{journal}{Nature}
  \textbf{\bibinfo{volume}{432}}, \bibinfo{pages}{81} (\bibinfo{year}{2004}).

\bibitem[{\citenamefont{Oulton et~al.}(2006)\citenamefont{Oulton, Greilich,
  Verbin, Cherbunin, Auer, Yakovlev, Bayer, Stavarache, Reuter, and
  Wieck}}]{oulton06}
\bibinfo{author}{\bibfnamefont{R.}~\bibnamefont{Oulton}},
  \bibinfo{author}{\bibfnamefont{A.}~\bibnamefont{Greilich}},
  \bibinfo{author}{\bibfnamefont{S.~Y.} \bibnamefont{Verbin}},
  \bibinfo{author}{\bibfnamefont{R.}~\bibnamefont{Cherbunin}},
  \bibinfo{author}{\bibfnamefont{T.}~\bibnamefont{Auer}},
  \bibinfo{author}{\bibfnamefont{D.~R.} \bibnamefont{Yakovlev}},
  \bibinfo{author}{\bibfnamefont{M.}~\bibnamefont{Bayer}},
  \bibinfo{author}{\bibfnamefont{V.}~\bibnamefont{Stavarache}},
  \bibinfo{author}{\bibfnamefont{D.}~\bibnamefont{Reuter}}, \bibnamefont{and}
  \bibinfo{author}{\bibfnamefont{A.}~\bibnamefont{Wieck}},
  \bibinfo{journal}{cond-mat} \textbf{\bibinfo{volume}{v3}},
  \bibinfo{pages}{0505446} (\bibinfo{year}{2006}).

\bibitem[{\citenamefont{Merkulov et~al.}(2002)\citenamefont{Merkulov, Efros,
  and Rosen}}]{Merkulov02}
\bibinfo{author}{\bibfnamefont{I.~A.} \bibnamefont{Merkulov}},
  \bibinfo{author}{\bibfnamefont{A.~L.} \bibnamefont{Efros}}, \bibnamefont{and}
  \bibinfo{author}{\bibfnamefont{M.}~\bibnamefont{Rosen}},
  \bibinfo{journal}{Phys.\ Rev. B} \textbf{\bibinfo{volume}{65}},
  \bibinfo{pages}{205309} (\bibinfo{year}{2002}).

\bibitem[{\citenamefont{Khaetskii et~al.}(2002)\citenamefont{Khaetskii, Loss,
  and Glazman}}]{Khaet1}
\bibinfo{author}{\bibfnamefont{A.}~\bibnamefont{Khaetskii}},
  \bibinfo{author}{\bibfnamefont{D.}~\bibnamefont{Loss}}, \bibnamefont{and}
  \bibinfo{author}{\bibfnamefont{L.}~\bibnamefont{Glazman}},
  \bibinfo{journal}{Phys.\ Rev. Lett.} \textbf{\bibinfo{volume}{88}},
  \bibinfo{pages}{186802} (\bibinfo{year}{2002}).

\bibitem[{\citenamefont{Semenov and Kim}(2003)}]{Seme1}
\bibinfo{author}{\bibfnamefont{Y.}~\bibnamefont{Semenov}} \bibnamefont{and}
  \bibinfo{author}{\bibfnamefont{K.~W.} \bibnamefont{Kim}},
  \bibinfo{journal}{Phys.\ Rev. B} \textbf{\bibinfo{volume}{67}},
  \bibinfo{pages}{73301} (\bibinfo{year}{2003}).

\bibitem[{\citenamefont{Braun et~al.}(2005)\citenamefont{Braun, Marie, Lombez,
  Urbaszek, Amand, Renucci, Kalevich, Kavokin, Krebs, Voisin et~al.}}]{pif05}
\bibinfo{author}{\bibfnamefont{P.-F.} \bibnamefont{Braun}},
  \bibinfo{author}{\bibfnamefont{X.}~\bibnamefont{Marie}},
  \bibinfo{author}{\bibfnamefont{L.}~\bibnamefont{Lombez}},
  \bibinfo{author}{\bibfnamefont{B.}~\bibnamefont{Urbaszek}},
  \bibinfo{author}{\bibfnamefont{T.}~\bibnamefont{Amand}},
  \bibinfo{author}{\bibfnamefont{P.}~\bibnamefont{Renucci}},
  \bibinfo{author}{\bibfnamefont{V.~K.} \bibnamefont{Kalevich}},
  \bibinfo{author}{\bibfnamefont{K.~V.} \bibnamefont{Kavokin}},
  \bibinfo{author}{\bibfnamefont{O.}~\bibnamefont{Krebs}},
  \bibinfo{author}{\bibfnamefont{P.}~\bibnamefont{Voisin}},
  \bibnamefont{et~al.}, \bibinfo{journal}{Phys.\ Rev. Lett.}
  \textbf{\bibinfo{volume}{94}}, \bibinfo{pages}{116601}
  (\bibinfo{year}{2005}).

\bibitem[{\citenamefont{Lampel}(1968)}]{Lampel68}
\bibinfo{author}{\bibfnamefont{G.}~\bibnamefont{Lampel}},
  \bibinfo{journal}{Phys.\ Rev. Lett.} \textbf{\bibinfo{volume}{20}},
  \bibinfo{pages}{491} (\bibinfo{year}{1968}).

\bibitem[{\citenamefont{Gammon et~al.}(2001)\citenamefont{Gammon, Efros,
  Kennedy, Rosen, Katzer, Park, Brown, Korenev, and Merkulov}}]{GammonPRL}
\bibinfo{author}{\bibfnamefont{D.}~\bibnamefont{Gammon}},
  \bibinfo{author}{\bibfnamefont{A.~L.} \bibnamefont{Efros}},
  \bibinfo{author}{\bibfnamefont{T.~A.} \bibnamefont{Kennedy}},
  \bibinfo{author}{\bibfnamefont{M.}~\bibnamefont{Rosen}},
  \bibinfo{author}{\bibfnamefont{D.~S.} \bibnamefont{Katzer}},
  \bibinfo{author}{\bibfnamefont{D.}~\bibnamefont{Park}},
  \bibinfo{author}{\bibfnamefont{S.~W.} \bibnamefont{Brown}},
  \bibinfo{author}{\bibfnamefont{V.~L.} \bibnamefont{Korenev}},
  \bibnamefont{and} \bibinfo{author}{\bibfnamefont{I.~A.}
  \bibnamefont{Merkulov}}, \bibinfo{journal}{Phys.\ Rev. Lett.}
  \textbf{\bibinfo{volume}{86}}, \bibinfo{pages}{5176} (\bibinfo{year}{2001}).

\bibitem[{\citenamefont{Bracker et~al.}(2005)\citenamefont{Bracker, Stinaff,
  Gammon, Ware, Tischler, Shabaev, Efros, Park, Gerschoni, Korenev
  et~al.}}]{Bracker1}
\bibinfo{author}{\bibfnamefont{A.}~\bibnamefont{Bracker}},
  \bibinfo{author}{\bibfnamefont{E.~A.} \bibnamefont{Stinaff}},
  \bibinfo{author}{\bibfnamefont{D.}~\bibnamefont{Gammon}},
  \bibinfo{author}{\bibfnamefont{M.~E.} \bibnamefont{Ware}},
  \bibinfo{author}{\bibfnamefont{J.~G.} \bibnamefont{Tischler}},
  \bibinfo{author}{\bibfnamefont{A.}~\bibnamefont{Shabaev}},
  \bibinfo{author}{\bibfnamefont{A.~L.} \bibnamefont{Efros}},
  \bibinfo{author}{\bibfnamefont{D.}~\bibnamefont{Park}},
  \bibinfo{author}{\bibfnamefont{D.}~\bibnamefont{Gerschoni}},
  \bibinfo{author}{\bibfnamefont{V.~L.} \bibnamefont{Korenev}},
  \bibnamefont{et~al.}, \bibinfo{journal}{Phys. Rev. Lett.}
  \textbf{\bibinfo{volume}{94}}, \bibinfo{pages}{047402}
  (\bibinfo{year}{2005}).

\bibitem[{opo()}]{opor}
\bibinfo{note}{\emph{Optical Orientation}, edited by F. Meier and B.
  Zakharchenya, Modern Porblems in Condensed Matter Sciences, Vol 8.
  (North-Holland, Amsterdam, 1984).}

\bibitem[{\citenamefont{Paget}(1982)}]{Paget82}
\bibinfo{author}{\bibfnamefont{D.}~\bibnamefont{Paget}},
  \bibinfo{journal}{Phys.\ Rev. B} \textbf{\bibinfo{volume}{25}},
  \bibinfo{pages}{4444} (\bibinfo{year}{1982}).

\bibitem[{\citenamefont{Imamoglu et~al.}(2003)\citenamefont{Imamoglu, Knill,
  Tian, and Zoller}}]{Ima2}
\bibinfo{author}{\bibfnamefont{A.}~\bibnamefont{Imamoglu}},
  \bibinfo{author}{\bibfnamefont{E.}~\bibnamefont{Knill}},
  \bibinfo{author}{\bibfnamefont{L.}~\bibnamefont{Tian}}, \bibnamefont{and}
  \bibinfo{author}{\bibfnamefont{P.}~\bibnamefont{Zoller}},
  \bibinfo{journal}{Phys.\ Rev. Lett.} \textbf{\bibinfo{volume}{91}},
  \bibinfo{pages}{17402} (\bibinfo{year}{2003}).

\bibitem[{\citenamefont{Gammon et~al.}(1997)\citenamefont{Gammon, Brown, Snow,
  Kennedy, Katzer, and Park}}]{Gammon1}
\bibinfo{author}{\bibfnamefont{D.}~\bibnamefont{Gammon}},
  \bibinfo{author}{\bibfnamefont{S.~W.} \bibnamefont{Brown}},
  \bibinfo{author}{\bibfnamefont{E.~S.} \bibnamefont{Snow}},
  \bibinfo{author}{\bibfnamefont{T.~A.} \bibnamefont{Kennedy}},
  \bibinfo{author}{\bibfnamefont{D.~S.} \bibnamefont{Katzer}},
  \bibnamefont{and} \bibinfo{author}{\bibfnamefont{D.}~\bibnamefont{Park}},
  \bibinfo{journal}{Science} \textbf{\bibinfo{volume}{277}},
  \bibinfo{pages}{87} (\bibinfo{year}{1997}).

\bibitem[{\citenamefont{Yokoi et~al.}(2005)\citenamefont{Yokoi, Adachi,
  Sasakura, Muto, Song, Usuki, and Hirose}}]{Yokoi}
\bibinfo{author}{\bibfnamefont{T.}~\bibnamefont{Yokoi}},
  \bibinfo{author}{\bibfnamefont{S.}~\bibnamefont{Adachi}},
  \bibinfo{author}{\bibfnamefont{H.}~\bibnamefont{Sasakura}},
  \bibinfo{author}{\bibfnamefont{S.}~\bibnamefont{Muto}},
  \bibinfo{author}{\bibfnamefont{H.}~\bibnamefont{Song}},
  \bibinfo{author}{\bibfnamefont{T.}~\bibnamefont{Usuki}}, \bibnamefont{and}
  \bibinfo{author}{\bibfnamefont{S.}~\bibnamefont{Hirose}},
  \bibinfo{journal}{Phys.\ Rev. B} \textbf{\bibinfo{volume}{71}},
  \bibinfo{pages}{R041307} (\bibinfo{year}{2005}).

\bibitem[{\citenamefont{Eble et~al.}(2006)\citenamefont{Eble, Krebs, Lemaitre,
  Kowalik, Kudelski, Voisin, Urbaszek, Marie, and Amand}}]{Eble}
\bibinfo{author}{\bibfnamefont{B.}~\bibnamefont{Eble}},
  \bibinfo{author}{\bibfnamefont{O.}~\bibnamefont{Krebs}},
  \bibinfo{author}{\bibfnamefont{A.}~\bibnamefont{Lemaitre}},
  \bibinfo{author}{\bibfnamefont{K.}~\bibnamefont{Kowalik}},
  \bibinfo{author}{\bibfnamefont{A.}~\bibnamefont{Kudelski}},
  \bibinfo{author}{\bibfnamefont{P.}~\bibnamefont{Voisin}},
  \bibinfo{author}{\bibfnamefont{B.}~\bibnamefont{Urbaszek}},
  \bibinfo{author}{\bibfnamefont{X.}~\bibnamefont{Marie}}, \bibnamefont{and}
  \bibinfo{author}{\bibfnamefont{T.}~\bibnamefont{Amand}},
  \bibinfo{journal}{Phys. Rev. B} \textbf{\bibinfo{volume}{74}},
  \bibinfo{pages}{R081306} (\bibinfo{year}{2006}).

\bibitem[{\citenamefont{Lai et~al.}(2006)\citenamefont{Lai, Maletinsky,
  Badolato, and Imamoglu}}]{Lai}
\bibinfo{author}{\bibfnamefont{C.~W.} \bibnamefont{Lai}},
  \bibinfo{author}{\bibfnamefont{P.}~\bibnamefont{Maletinsky}},
  \bibinfo{author}{\bibfnamefont{A.}~\bibnamefont{Badolato}}, \bibnamefont{and}
  \bibinfo{author}{\bibfnamefont{A.}~\bibnamefont{Imamoglu}},
  \bibinfo{journal}{Phys.\ Rev. Lett.} \textbf{\bibinfo{volume}{96}},
  \bibinfo{pages}{167403} (\bibinfo{year}{2006}).

\bibitem[{\citenamefont{Dzhioev et~al.}(1999)\citenamefont{Dzhioev,
  Zakharchenya, Korenev, and Lazarev}}]{Dzhioev1}
\bibinfo{author}{\bibfnamefont{R.}~\bibnamefont{Dzhioev}},
  \bibinfo{author}{\bibfnamefont{B.~P.} \bibnamefont{Zakharchenya}},
  \bibinfo{author}{\bibfnamefont{V.~L.} \bibnamefont{Korenev}},
  \bibnamefont{and} \bibinfo{author}{\bibfnamefont{M.~V.}
  \bibnamefont{Lazarev}}, \bibinfo{journal}{Phys.\ Sol. State}
  \textbf{\bibinfo{volume}{41}}, \bibinfo{pages}{2014} (\bibinfo{year}{1999}).

\bibitem[{Ign()}]{Igna1}
\bibinfo{note}{I. V. Ignatiev, I. Ya. Gerlovin, S. Yu. Verbin, W. Maruyama and
  Y. Masumoto, 13th Int. Symp. Nanostructures:Physics and Technology p. 47
  (2005)}.

\bibitem[{\citenamefont{Lemaitre et~al.}(2004)\citenamefont{Lemaitre,
  Patriarche, and Glas}}]{AL2004}
\bibinfo{author}{\bibfnamefont{A.}~\bibnamefont{Lemaitre}},
  \bibinfo{author}{\bibfnamefont{G.}~\bibnamefont{Patriarche}},
  \bibnamefont{and} \bibinfo{author}{\bibfnamefont{F.}~\bibnamefont{Glas}},
  \bibinfo{journal}{Appl. Phys. Lett.} \textbf{\bibinfo{volume}{85}},
  \bibinfo{pages}{3717} (\bibinfo{year}{2004}).

\bibitem[{\citenamefont{Vasanelli et~al.}(2002)\citenamefont{Vasanelli,
  Ferreira, and Bastard}}]{Vasa2002}
\bibinfo{author}{\bibfnamefont{A.}~\bibnamefont{Vasanelli}},
  \bibinfo{author}{\bibfnamefont{R.}~\bibnamefont{Ferreira}}, \bibnamefont{and}
  \bibinfo{author}{\bibfnamefont{G.}~\bibnamefont{Bastard}},
  \bibinfo{journal}{Phys. Rev. Lett.} \textbf{\bibinfo{volume}{89}},
  \bibinfo{pages}{216804} (\bibinfo{year}{2002}).

\bibitem[{\citenamefont{Akimov et~al.}(2002)\citenamefont{Akimov, Hundt,
  Flissikowski, and Henneberger}}]{Aki2002}
\bibinfo{author}{\bibfnamefont{I.}~\bibnamefont{Akimov}},
  \bibinfo{author}{\bibfnamefont{A.}~\bibnamefont{Hundt}},
  \bibinfo{author}{\bibfnamefont{T.}~\bibnamefont{Flissikowski}},
  \bibnamefont{and}
  \bibinfo{author}{\bibfnamefont{F.}~\bibnamefont{Henneberger}},
  \bibinfo{journal}{Appl.\ Phys. Lett.} \textbf{\bibinfo{volume}{81}},
  \bibinfo{pages}{4730} (\bibinfo{year}{2002}).

\bibitem[{\citenamefont{Bayer et~al.}(1999)\citenamefont{Bayer, Kuther,
  Forchel, Gorbunov, Timofeev, Schäfer, Reithmaier, Reinecke, and Walck}}]{Mb2}
\bibinfo{author}{\bibfnamefont{M.}~\bibnamefont{Bayer}},
  \bibinfo{author}{\bibfnamefont{A.}~\bibnamefont{Kuther}},
  \bibinfo{author}{\bibfnamefont{A.}~\bibnamefont{Forchel}},
  \bibinfo{author}{\bibfnamefont{A.}~\bibnamefont{Gorbunov}},
  \bibinfo{author}{\bibfnamefont{V.~B.} \bibnamefont{Timofeev}},
  \bibinfo{author}{\bibfnamefont{F.}~\bibnamefont{Schäfer}},
  \bibinfo{author}{\bibfnamefont{J.~P.} \bibnamefont{Reithmaier}},
  \bibinfo{author}{\bibfnamefont{T.~L.} \bibnamefont{Reinecke}},
  \bibnamefont{and} \bibinfo{author}{\bibfnamefont{S.~N.} \bibnamefont{Walck}},
  \bibinfo{journal}{Phys. Rev. Lett.} \textbf{\bibinfo{volume}{82}},
  \bibinfo{pages}{1748} (\bibinfo{year}{1999}).

\bibitem[{\citenamefont{M.Senes et~al.}(2005)\citenamefont{M.Senes, Urbaszek,
  Marie, Amand, Tribollet, Bernardot, Testelin, Chamarro, and Gerard}}]{Senes2}
\bibinfo{author}{\bibnamefont{M.Senes}},
  \bibinfo{author}{\bibfnamefont{B.}~\bibnamefont{Urbaszek}},
  \bibinfo{author}{\bibfnamefont{X.}~\bibnamefont{Marie}},
  \bibinfo{author}{\bibfnamefont{T.}~\bibnamefont{Amand}},
  \bibinfo{author}{\bibfnamefont{J.}~\bibnamefont{Tribollet}},
  \bibinfo{author}{\bibfnamefont{F.}~\bibnamefont{Bernardot}},
  \bibinfo{author}{\bibfnamefont{C.}~\bibnamefont{Testelin}},
  \bibinfo{author}{\bibfnamefont{M.}~\bibnamefont{Chamarro}}, \bibnamefont{and}
  \bibinfo{author}{\bibfnamefont{J.-M.} \bibnamefont{Gerard}},
  \bibinfo{journal}{Phys. Rev. B} \textbf{\bibinfo{volume}{71}},
  \bibinfo{pages}{115334} (\bibinfo{year}{2005}).

\bibitem[{\citenamefont{Tartakovskii et~al.}(2004)\citenamefont{Tartakovskii,
  Cahill, Makhonin, Whittaker, Wells, Fox, Mowbray, Skolnick, Groom, Steer
  et~al.}}]{Tarta2}
\bibinfo{author}{\bibfnamefont{A.}~\bibnamefont{Tartakovskii}},
  \bibinfo{author}{\bibfnamefont{J.}~\bibnamefont{Cahill}},
  \bibinfo{author}{\bibfnamefont{M.~N.} \bibnamefont{Makhonin}},
  \bibinfo{author}{\bibfnamefont{D.~M.} \bibnamefont{Whittaker}},
  \bibinfo{author}{\bibfnamefont{J.-P.~R.} \bibnamefont{Wells}},
  \bibinfo{author}{\bibfnamefont{A.~M.} \bibnamefont{Fox}},
  \bibinfo{author}{\bibfnamefont{D.~J.} \bibnamefont{Mowbray}},
  \bibinfo{author}{\bibfnamefont{M.~S.} \bibnamefont{Skolnick}},
  \bibinfo{author}{\bibfnamefont{K.~M.} \bibnamefont{Groom}},
  \bibinfo{author}{\bibfnamefont{M.~J.} \bibnamefont{Steer}},
  \bibnamefont{et~al.}, \bibinfo{journal}{Phys. Rev. Lett.}
  \textbf{\bibinfo{volume}{93}}, \bibinfo{pages}{057401}
  (\bibinfo{year}{2004}).

\bibitem[{\citenamefont{Langbein et~al.}(2004)\citenamefont{Langbein, Borri,
  Woggon, Stavarache, Reuter, and Wieck}}]{Lang2004}
\bibinfo{author}{\bibfnamefont{W.}~\bibnamefont{Langbein}},
  \bibinfo{author}{\bibfnamefont{P.}~\bibnamefont{Borri}},
  \bibinfo{author}{\bibfnamefont{U.}~\bibnamefont{Woggon}},
  \bibinfo{author}{\bibfnamefont{V.}~\bibnamefont{Stavarache}},
  \bibinfo{author}{\bibfnamefont{D.}~\bibnamefont{Reuter}}, \bibnamefont{and}
  \bibinfo{author}{\bibfnamefont{A.~D.} \bibnamefont{Wieck}},
  \bibinfo{journal}{Phys.\ Rev. B.} \textbf{\bibinfo{volume}{69}},
  \bibinfo{pages}{R161301} (\bibinfo{year}{2004}).

\bibitem[{Edi()}]{Ediger06}
\bibinfo{note}{M. Ediger, G. Bester, B.D. Gerardot, A. Badolato, P. M. Petroff,
  K. Karrai, A. Zunger and R. J. Warburton (to be published)}.

\bibitem[{\citenamefont{Gryncharova and Perel}(1977)}]{Gr1}
\bibinfo{author}{\bibfnamefont{E.~I.} \bibnamefont{Gryncharova}}
  \bibnamefont{and} \bibinfo{author}{\bibfnamefont{V.~I.} \bibnamefont{Perel}},
  \bibinfo{journal}{Sov. Phys. Semicond.} \textbf{\bibinfo{volume}{11}},
  \bibinfo{pages}{997} (\bibinfo{year}{1977}).

\bibitem[{Abr()}]{Abra}
\bibinfo{note}{A. Abragam, \emph{Principles of Nuclear Magnetism}, (Oxford
  University Press, 1961).}

\bibitem[{Aki()}]{Aki2006}
\bibinfo{note}{I.A. Akimov,D. H. Feng and F. Henneberger, Phys.\ Rev. Lett. (in
  press)}.

\bibitem[{Tex()}]{Text8}
\bibinfo{note}{The correlation time $\tau_c$ can be considerably smaller than
  the radiative lifetime $\tau_{rad}$ as the doping hole tunnels in and out of
  the dot.}

\bibitem[{han()}]{hand}
\bibinfo{note}{\emph{Handbook of Chemistry and Physics}, edited by R.C. Weast
  (The Chemical Rubber Co, Ohio 1968).}

\bibitem[{\citenamefont{He et~al.}(2004)\citenamefont{He, Bester, and
  Zunger}}]{Zunger2004}
\bibinfo{author}{\bibfnamefont{L.}~\bibnamefont{He}},
  \bibinfo{author}{\bibfnamefont{G.}~\bibnamefont{Bester}}, \bibnamefont{and}
  \bibinfo{author}{\bibfnamefont{A.}~\bibnamefont{Zunger}},
  \bibinfo{journal}{Phys.\ Rev. B} \textbf{\bibinfo{volume}{70}},
  \bibinfo{pages}{235316} (\bibinfo{year}{2004}).

\bibitem[{\citenamefont{Deng and Hu}(2005)}]{Deng}
\bibinfo{author}{\bibfnamefont{C.}~\bibnamefont{Deng}} \bibnamefont{and}
  \bibinfo{author}{\bibfnamefont{X.}~\bibnamefont{Hu}}, \bibinfo{journal}{Phys.
  Rev. B} \textbf{\bibinfo{volume}{71}}, \bibinfo{pages}{033307}
  (\bibinfo{year}{2005}).

\bibitem[{\citenamefont{Artemova et~al.}(1991)\citenamefont{Artemova,
  Galationov, Kalevich, Korenev, Merkulov, and Silbergleit}}]{Arte}
\bibinfo{author}{\bibfnamefont{E.~S.} \bibnamefont{Artemova}},
  \bibinfo{author}{\bibfnamefont{E.~V.} \bibnamefont{Galationov}},
  \bibinfo{author}{\bibfnamefont{V.~K.} \bibnamefont{Kalevich}},
  \bibinfo{author}{\bibfnamefont{V.~L.} \bibnamefont{Korenev}},
  \bibinfo{author}{\bibfnamefont{I.~A.} \bibnamefont{Merkulov}},
  \bibnamefont{and} \bibinfo{author}{\bibfnamefont{A.~S.}
  \bibnamefont{Silbergleit}}, \bibinfo{journal}{Nonlinearity}
  \textbf{\bibinfo{volume}{4}}, \bibinfo{pages}{49} (\bibinfo{year}{1991}).

\bibitem[{Tar()}]{Tarta1}
\bibinfo{note}{A.I. Tartakovskii et al, cond-mat/0609371}.

\bibitem[{Ima()}]{Imapriv}
\bibinfo{note}{P. Maletinsky et al, cond-mat/0609291}.

\bibitem[{\citenamefont{Kalevich and Korenev}(1992)}]{Kal92}
\bibinfo{author}{\bibfnamefont{V.}~\bibnamefont{Kalevich}} \bibnamefont{and}
  \bibinfo{author}{\bibfnamefont{V.}~\bibnamefont{Korenev}},
  \bibinfo{journal}{JETP Lett.} \textbf{\bibinfo{volume}{56}},
  \bibinfo{pages}{253} (\bibinfo{year}{1992}).

\bibitem[{\citenamefont{Sanada et~al.}(2004)\citenamefont{Sanada, Matsuzaka,
  Morita, Hu, Ohno, and Ohno}}]{San2004}
\bibinfo{author}{\bibfnamefont{H.}~\bibnamefont{Sanada}},
  \bibinfo{author}{\bibfnamefont{S.}~\bibnamefont{Matsuzaka}},
  \bibinfo{author}{\bibfnamefont{K.}~\bibnamefont{Morita}},
  \bibinfo{author}{\bibfnamefont{C.}~\bibnamefont{Hu}},
  \bibinfo{author}{\bibfnamefont{Y.}~\bibnamefont{Ohno}}, \bibnamefont{and}
  \bibinfo{author}{\bibfnamefont{H.}~\bibnamefont{Ohno}},
  \bibinfo{journal}{Phys. Rev. B} \textbf{\bibinfo{volume}{68}},
  \bibinfo{pages}{R241303} (\bibinfo{year}{2004}).


\end{thebibliography}

\end{document}